\documentclass[aps,prb,reprint,superscriptaddress]{revtex4-1}
\usepackage{graphicx}
\usepackage{pstricks}
\usepackage{amsmath}
\usepackage{dcolumn}
\newcommand{\dfe}{$d_{\rm Fe-Fe}$}
\newcommand{\dse}{$d_{\rm Fe-Se}$}

\newcommand{\dfedse}{$(d_{\rm Fe-Fe},d_{\rm Fe-Se})$}

\begin{document}
\title{Basic electronic properties of iron selenide under variation of structural parameters}

\author{Daniel Guterding} 
\altaffiliation[Present address: ]{Lucht Probst Associates, Gro{\ss}e Gallusstra{\ss}e 9, 60311 Frankfurt am Main, Germany}
\email{daniel.guterding@gmail.com}
\affiliation{Institut f\"ur Theoretische Physik, Goethe-Universit\"at Frankfurt, 
Max-von-Laue-Stra{\ss}e 1, 60438 Frankfurt am Main, Germany}
\author{Harald O. Jeschke}
\affiliation{Research Institute for 
Interdisciplinary Science, Okayama University, Okayama 700-8530, Japan}
\author{Roser Valent\'i}
\affiliation{Institut f\"ur Theoretische Physik, Goethe-Universit\"at Frankfurt,
Max-von-Laue-Stra{\ss}e 1, 60438 Frankfurt am Main, Germany}

\begin{abstract}
Since the discovery of high-temperature superconductivity in the thin-film
FeSe/SrTiO$_3$ system, iron selenide and its derivates have been 
intensively scrutinized. Using \textit{ab initio} density functional theory calculations we
review the electronic structures that could be realized in iron-selenide
if the structural parameters could be tuned at liberty. We calculate the
momentum-dependence of the susceptibility and investigate the symmetry of
electron pairing within the random phase approximation. Both the
susceptibility and the symmetry of electron pairing depend on the structural
parameters in a nontrivial way.  These results are consistent with the known
experimental behavior of binary iron chalcogenides and, at the same time, reveal two
promising ways of tuning superconducting transition temperatures in these
materials. On the one hand by expanding the iron lattice of FeSe at constant
 iron-selenium distance and, on the other hand, by
 increasing the iron-selenium distance with unchanged iron lattice.
\end{abstract}


\maketitle

\section{Introduction}

Since the discovery of iron based superconductors in 2008, this field
 has matured, with many efforts going
into tuning properties of the materials towards higher transition
temperatures, higher critical fields, better crystal properties, or
less rare constituents.  The iron chalcogenides with its primary
exponent FeSe~\cite{Hsu2008,Coldea2017} have recently attracted intense
scrutiny. On the one hand, there are experimental facts like the large
nematic region~\cite{boehmer2016}
 that continue to trigger 
theoretical efforts~\cite{glasbrenner2015,wang2015,chubukov2016,yamakawa2016,fanfarillo2016,scherer2017} in the
hope of improving and unifying our understanding of iron based
superconductors in general.  On the other hand, special types of
tuning like growth on substrates~\cite{Liu2012,Tang2013,Ge2014}, doping~\cite{Watson2015,Coldea2016,Matsuura2017,Lai2015, Ying2016, Tresca2017}
intercalation~\cite{Burrard-Lucas2013,Sedlmaier2014,Dong2014,Lynn2015,Sun2015,Lu2015,Pachmayr2015,Hayashi2015},
and the family of alkali iron chalcogenides~\cite{Guo2010}  make FeSe
and its derivates especially rich. There have been several attempts to
identify crucial tuning properties in FeSe, in particular the Se
height~\cite{Noji2014,Hosono2016} or the doping level. 

\begin{figure}[b]
  \includegraphics[width=\linewidth]{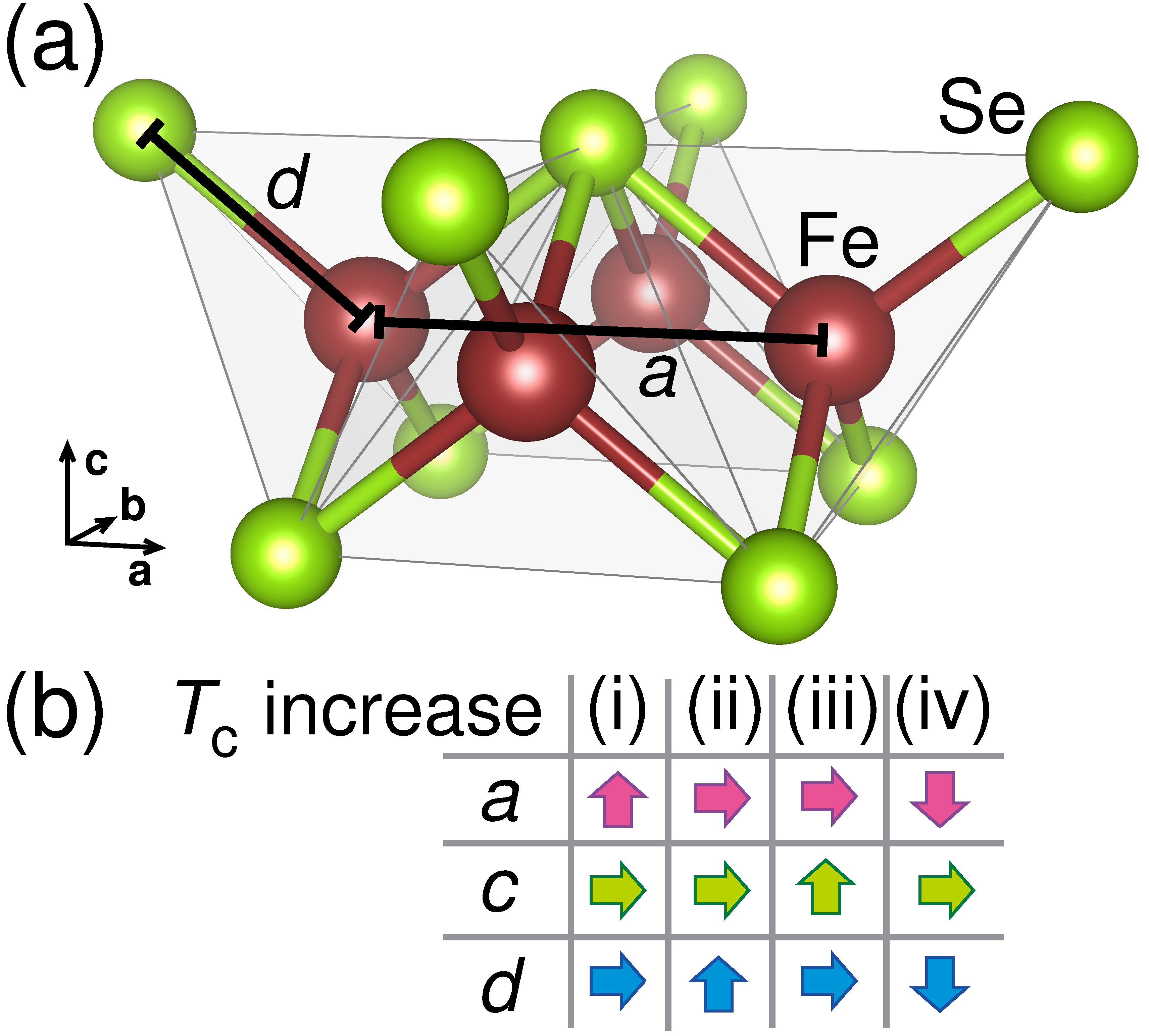}\caption{(Color
    online) Crystal structure of FeSe (top) and summary of the four
    ways to increase the superconducting transition temperature $T_c$
    (bottom) by changing lattice parameters $a$, $c$ and $d$, as
    described in the main text. Arrows indicate the change in lattice
    parameter. An up arrow represents enlargement, a down arrow
    represents shrinking, while a horizontal arrow represents keeping
    the parameter unchanged compared to the experimental pristine
    FeSe.}
\label{fig:introfigure}
\end{figure}

The tetragonal
FeSe crystal structure is intriguingly simple; it is fully specified
by $a$ and $c$ lattice parameters and the Se $z$ coordinate. However,
focusing more closely on a single layer of FeSe, it is clear that the
electronic structure is controlled by only two geometric parameters,
the Fe-Fe distance {\dfe} which is identical to the $a$ lattice
parameter divided by $\sqrt{2}$, and the Fe-Se distance {\dse}.  Experimentally, a large
range of these two distances {\dfe} and {\dse} can be realized using
pressure~\cite{Medvedev2009,Terashima2015}, strain~\cite{Wang2009},
substrates~\cite{Nie2009,Hiramatsu2012,Liu2012},
intercalation~\cite{Burrard-Lucas2013,Sedlmaier2014,Dong2014,Lynn2015,Sun2015,Lu2015,Pachmayr2015},
chalcogenide substitution~\cite{Watson2015,Coldea2016,Matsuura2017,Lai2015, Ying2016, Tresca2017}
and charge doping~\cite{Lei2016}. 

\begin{figure*}[t]
  \includegraphics[width=\linewidth]{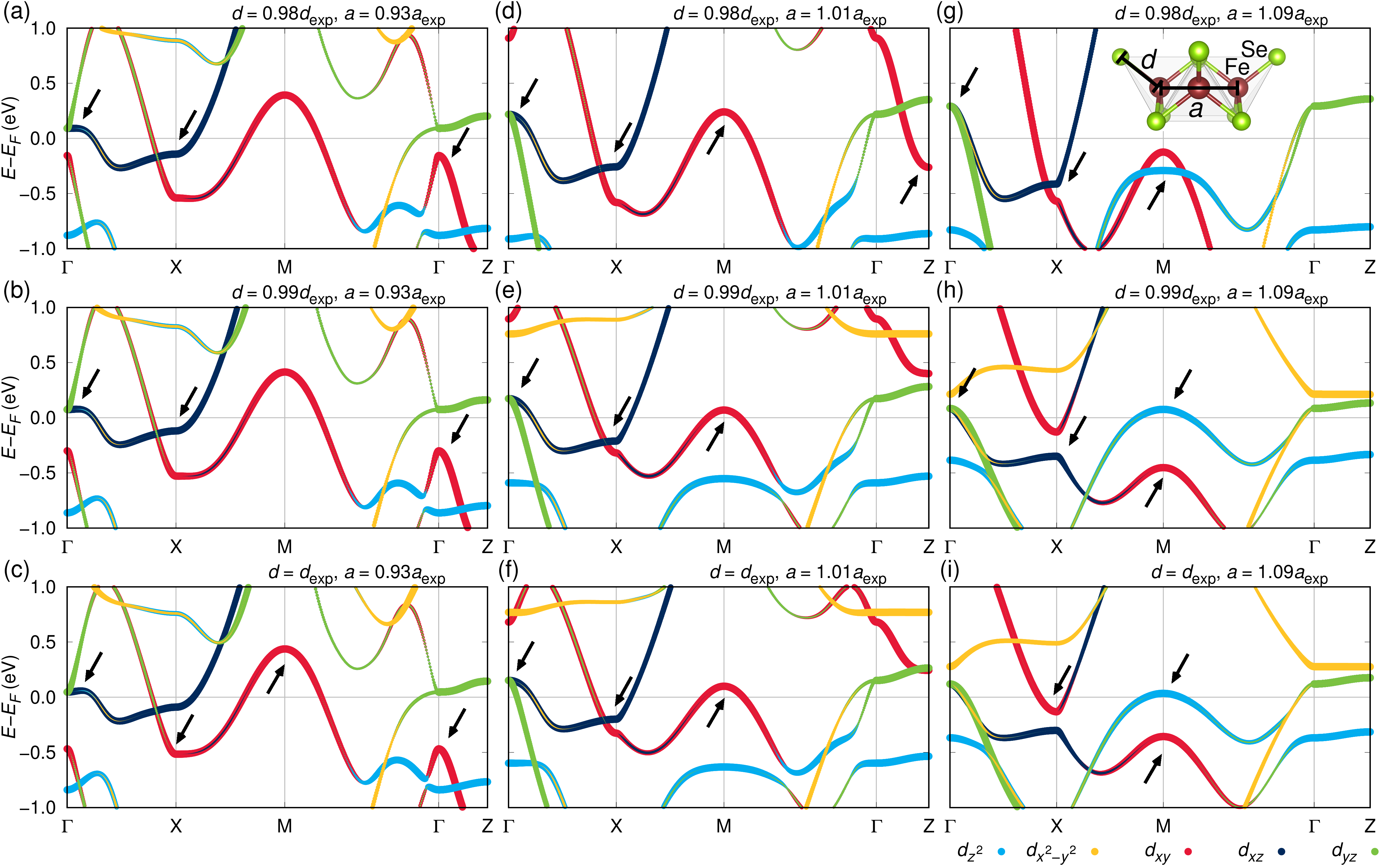}\caption{(Color
    online) Electronic bandstructure with orbital weights indicated by
    colors for varied iron-selenium distance {\dse} and lattice
    parameter $a$. (a)-(c) $a=3.5~{\rm \AA}=0.93a_{\rm exp}$, (d)-(f)
    $a=3.8~{\rm \AA}=1.01a_{\rm exp}$ and (g)-(i) $a=4.1~{\rm
      \AA}=1.09a_{\rm exp}$. First row (a), (d), (g) {\dse}$=2.35~{\rm
      \AA}=0.98d_{\rm exp}$, second row (b), (e), (h)
    {\dse}$=2.37~{\rm \AA}=0.99d_{\rm exp}$ and third row (c), (f),
    (i) {\dse}$ = 2.393~{\rm \AA}\equiv d_{\rm exp}$. Arrows mark
    points where changes in the bandstructure happen as a function of
    parameters.}
\label{fig:bandsoverview}
\end{figure*}

In this contribution, we intend to
scan the possible electronic structures that could be realized if
{\dfe} and {\dse} could be tuned at liberty. This is done with the
hope of identifying promising directions in the {\dfedse} parameter
space which could then be targeted by material design efforts. For the
present study, we intentionally leave out the other significant tuning
axis which is the filling of the Fe bands; we fix it to the charge of
the neutral FeSe layer. We then survey as function of {\dfedse} the
variability of the FeSe electronic structure, Fermi surface, essential
tight binding parameters, magnetic susceptibility, as well as pairing
symmetry and strength. We note that in this study we will neither handle
nematicity nor enter into the discussion of correlation 
effects~\cite{aichhorn,watson2017,aichhorn2,skornyakov2017}, but we will
rather concentrate on the basic trends of the electronic properties.

\begin{figure*}[t]
  \includegraphics[width=\linewidth]{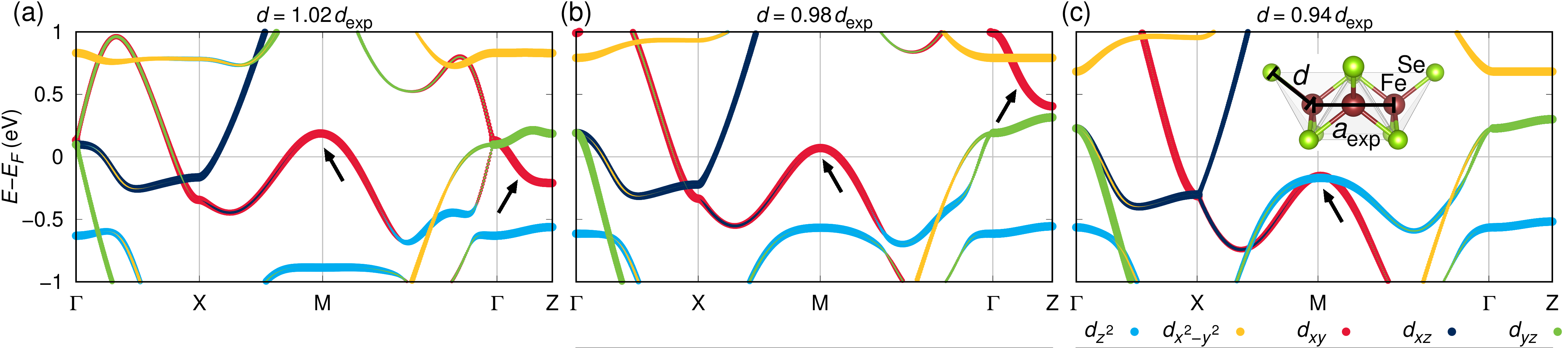}\caption{(Color
    online) Electronic bandstructure with orbital weights indicated by
    colors at constant lattice parameter $a=a_{\rm exp}$ and varied
    iron-selenium distance $d$. (a)
    {\dse}=$2.447~{\rm \AA}=1.02d_{\rm exp}$, (b)
    {\dse}=$2.348~{\rm \AA}=0.98d_{\rm exp}$ and (c)
    {\dse}=$2.256~{\rm \AA}=0.94d_{\rm exp}$. Arrows mark points
    where changes in the bandstructure happen as a function of
    parameters.}
\label{fig:bandsheightcontrol}
\end{figure*}

\begin{figure*}[t]
  \includegraphics[width=0.85\linewidth]{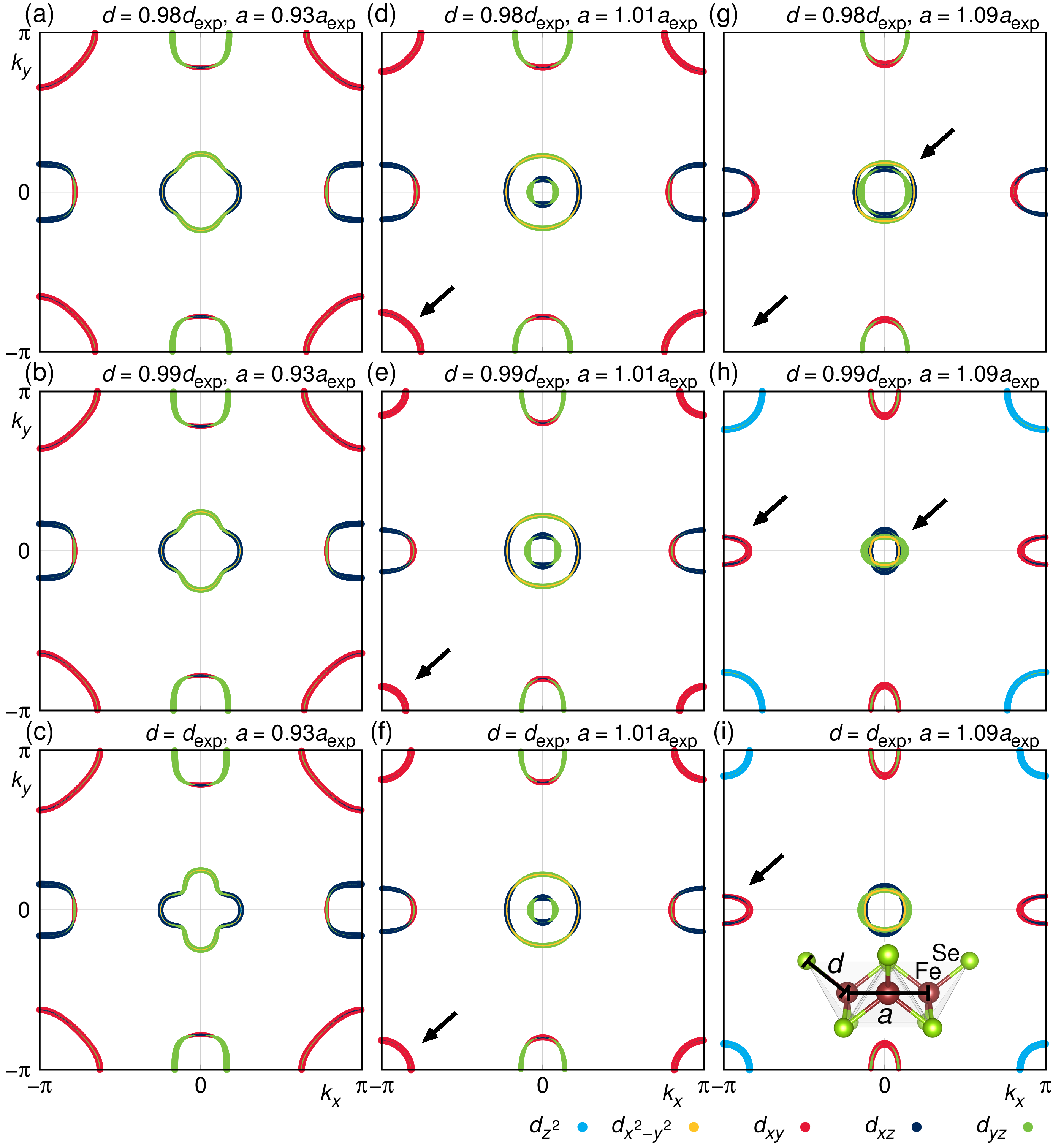}\caption{(Color
    online) Fermi surface with orbital weights indicated by colors for
    varied iron-selenium distance {\dse} and lattice parameter
    $a$. (a)-(c) $a=3.5~{\rm \AA}=0.93a_{\rm exp}$, (d)-(f)
    $a=3.8~{\rm \AA}=1.01a_{\rm exp}$ and (g)-(i) $a=4.1~{\rm
      \AA}=1.09a_{\rm exp}$. First row (a), (d), (g) {\dse}$=2.35~{\rm
      \AA}=0.98d_{\rm exp}$, second row (b), (e), (h)
    {\dse}$=2.37~{\rm \AA}=0.99d_{\rm exp}$ and third row (c), (f),
    (i) {\dse}$=2.393~{\rm \AA}\equiv d_{\rm exp}$. Arrows mark points
    where changes in the Fermi surface happen as a function of
    parameters.}
\label{fig:fsoverview}
\end{figure*}

\begin{figure*}[t]
  \includegraphics[width=0.85\linewidth]{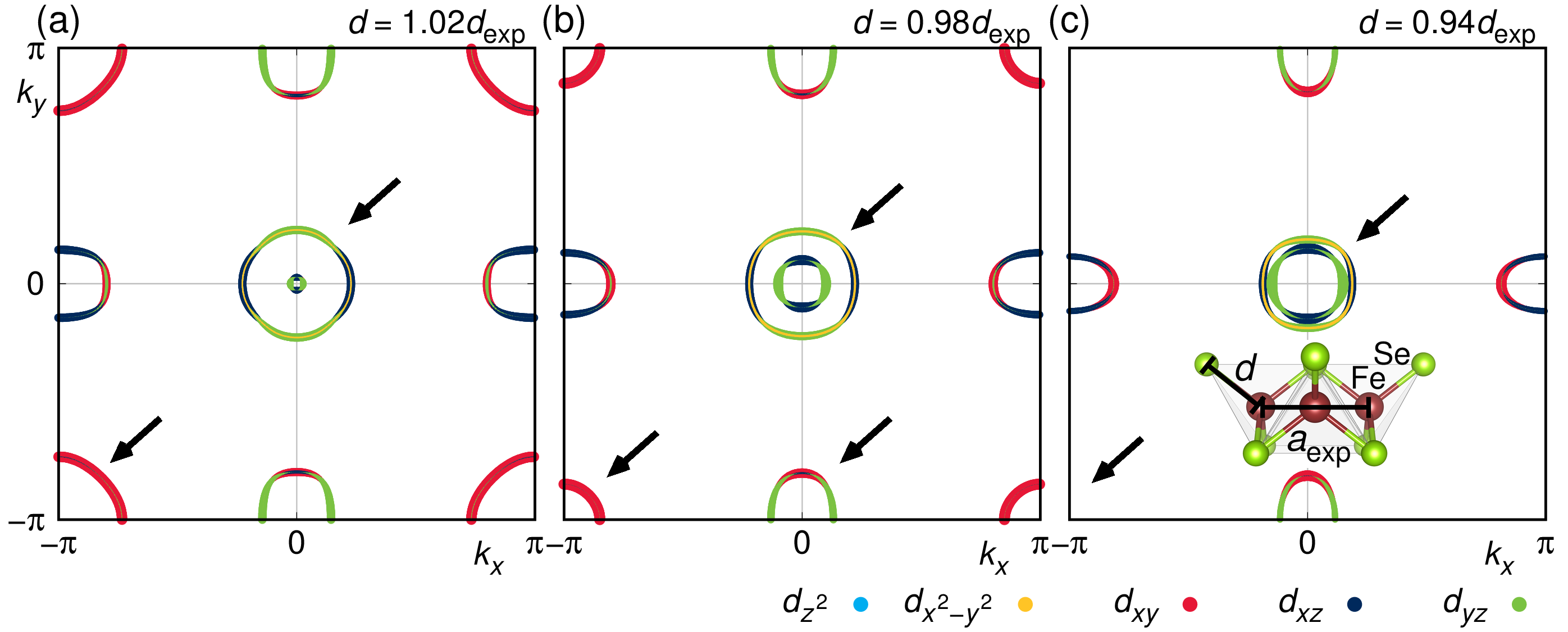}\caption{(Color
    online) Fermi surface with orbital weights indicated by colors at
    constant lattice parameter $a=a_{\rm exp}$ and varied
    iron-selenium distance $d$. (a)
    {\dse}=$2.447~{\rm \AA}=1.02d_{\rm exp}$, (b)
    {\dse}=$2.348~{\rm \AA}=0.98d_{\rm exp}$ and (c)
    {\dse}=$2.256~{\rm \AA}=0.94d_{\rm exp}$. Arrows mark points
    where changes in the Fermi surface happen as a function of
    parameters.}
\label{fig:fsheightcontrol}
\end{figure*}

Our main observations are: (i) Expansion of the
Fe square lattice parameter at constant Fe-Se distance {\dse} should
enhance $T_c$ up to a point before a switch of superconducting order
parameter from $s_\pm$ to $d_{x^2-y^2}$ occurs.  (ii) Alternatively,
increasing the Fe-Se
distance at constant Fe square lattice parameter $a$ should also enhance
$T_c$.  Furthermore, (iii) increasing the $c$ lattice parameter at constant
experimental {\dfe} and {\dse} distances (i.e. increasing the van der
Waals gap) only slightly increases $T_c$, while  (iv) compression of the Fe-Fe
square lattice at slightly compressed Fe-Se distance significantly enhances
$T_c$.  Observation (iii) is essentially known from charge neutral
FeSe intercalatates and observation (iv) is very consistent with the well
known pressure enhancement of $T_c$. However, observations (i) and
(ii) could lead to new design ideas. A summary of our findings is given in Fig.~\ref{fig:introfigure}.

\section{Methods and Models}
\subsection{\textit{Ab initio} calculations and model construction}
As a starting point for our calculations we use the $^{56}$FeSe$_{1-x}$ 
structure in space group P4/nmm (No.~129) obtained at $T=250~{\rm K}$ in 
Ref.~\onlinecite{Khasanov2010}. The structural parameters are $a = 
3.76988~\text{\AA}\equiv a_{\rm exp}$, $c = 5.51637~\text{\AA}\equiv c_{\rm 
exp}$ and $d_{\rm Fe-Se} = 2.393~\text{\AA}\equiv d_{\rm Fe-Se~exp}$. 

To investigate the dependence of the electronic structure on these parameters, 
we modify the crystal structures manually and calculate the electronic 
bandstructure using density functional theory (DFT) within the full-potential local 
orbital (FPLO)~\cite{FPLOmethod} basis independently for each case. We use the 
generalized gradient approximation~\cite{PerdewBurkeErnzerhof} for the 
exchange-correlation functional. All calculations were converged on $20 \times 
20 \times 20$ $\mathbf k$-point grids. 

Tight-binding models were constructed using projective Wannier 
functions~\cite{FPLOtightbinding}. The energy window chosen for the projection 
spans approximately from -2.5~eV to 2.0~eV. We include all Fe $3d$ states, which 
yields a ten-orbital model for the electronic structure, since the 
crystallographic unit cell contains two formula units of FeSe. Using the 
recently developed glide reflection unfolding technique~\cite{TomicUnfolding}, 
these ten-orbital models are reduced to five-orbital models for an effective 
one-iron unit cell. The calculated hopping parameters are denoted as $t_{ij}^{s 
p}$, where $i$ and $j$ are lattice site indices and indices $s$ and $p$ identify 
the orbitals.

\begin{equation}
\begin{split}
H_0 =& - \sum\limits_{i,j,s, p, \sigma}
t_{ij}^{s p} c^\dagger_{i s \sigma} c^{\,}_{j p \sigma}
\end{split}
\label{eq:kinetichamiltonian}
\end{equation}
The five-orbital models constitute the kinetic part of the full model 
Hamiltonian under investigation. Since the bandstructure calculated from DFT and 
the subsequently obtained tight-binding models are virtually identical in the 
energy window of interest, we use from now on only the tight-binding 
representation.

\begin{figure*}[t]
  \includegraphics[width=\linewidth]{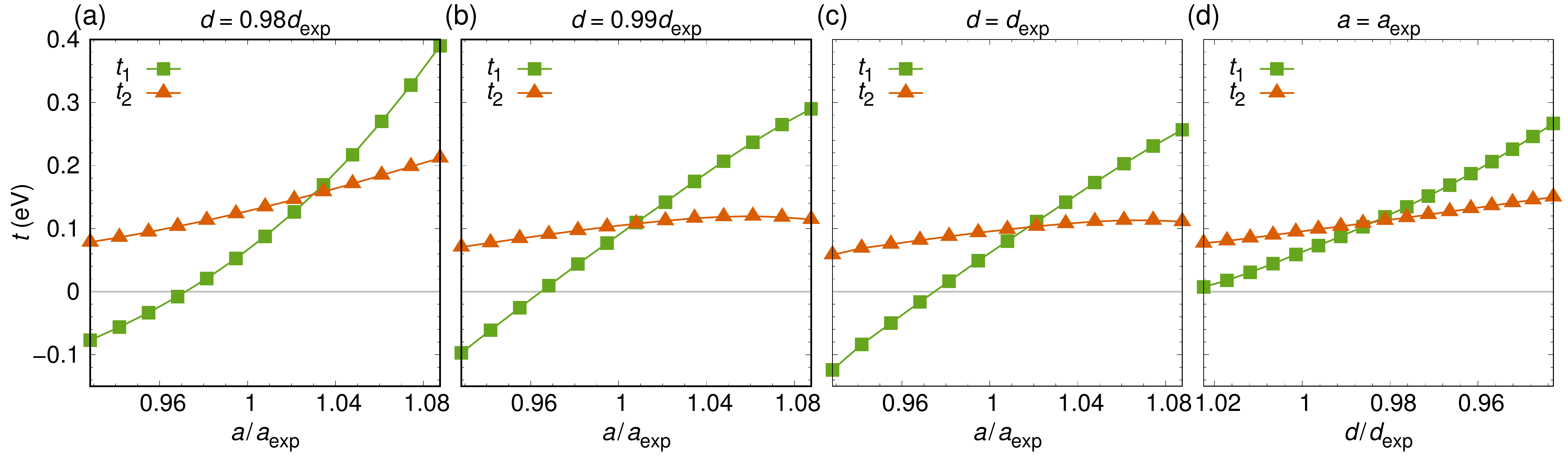}\caption{(Color
    online) Nearest ($t_1$) and next-nearest ($t_2$) neighbor hopping
    parameters in the Fe $3d_{xy}$ orbitals (a) as function of lattice
    parameters $a$ at fixed iron-selenium distance (a)
    {\dse}$=2.35~{\rm \AA}=0.98d_{\rm exp}$, (b)
    {\dse}$=2.37~{\rm \AA}=0.99d_{\rm exp}$ and (c)
    {\dse}$=2.393~{\rm \AA}\equiv d_{\rm exp}$. (d) The hoppings
    at fixed lattice parameter $a=3.76988~\text{\AA} \equiv a_{\rm
      exp}$ and varied iron-selenium distance {\dse}.}
\label{fig:tbparameters}
\end{figure*}

\subsection{Susceptibility and pairing calculations}
We assume that the interaction part of the full model Hamiltonian is given by the multi-orbital Hubbard interaction. Together with the kinetic part determined from the Wannier function calculation, the full Hamiltonian is given by

\begin{equation}
\begin{split}
H =& H_0 + H_{\rm int} \\ =& - \sum\limits_{i,j,s,p, \sigma}
t_{ij}^{s p} c^\dagger_{i s \sigma} c^{\,}_{j p \sigma} + U \sum\limits_{i,l} 
n_{i l \uparrow} n_{i l \downarrow} \\
& + \frac{U'}{2} \sum\limits_{i,s,p \neq s} n_{i s} n_{i p}
- \frac{J}{2}\sum\limits_{i,s,p\neq s}
\mathbf S_{i s} \cdot \mathbf S_{i p} \\
& + \frac{J^\prime}{2} \sum \limits_{i,s,p \neq s, \sigma} 
c^\dagger_{i s \sigma} c^\dagger_{i s \bar \sigma}
c^{\,}_{i p \bar \sigma} c^{\,}_{i p \sigma}
.
\end{split}
\label{eq:multiorbitalhubbard}
\end{equation}

The number operator is given by $n_{is\sigma} = c^\dagger_{i s\sigma} c^{\,}_{i s\sigma}$ and $\mathbf S_s = \frac{1}{2}\sum_{\mathbf \alpha \beta} c^\dagger_{s \alpha} \boldsymbol \sigma_{\alpha \beta} c^{\,}_{s \beta}$ is a spin operator with orbital index denoted by $s$ and the spins of electrons denoted by $\alpha$ and $\beta$. The interaction parameters are the intra-orbital Coulomb repulsion ($U$), the inter-orbital Coulomb repulsion ($U'$), the Hund's rule coupling ($J$) and the pair-hopping term ($J^\prime$). For the values of these parameters we make an assumption that is consistent with the existing literature: $U = 1.25~{\rm eV}$, $U' = U/2$, $J=J^\prime=U/4$.

Our aim is now to calculate the two-particle pairing vertex within the random phase approximation (RPA)~\cite{Graser2009, Altmeyer2016, Guterding2017}. We first calculate the non-interacting susceptibility $\chi_{st}^{pq} (\mathbf q)$ as a function of momentum, but at vanishing excitation frequency (static limit).

\begin{equation}
\begin{aligned}
\chi_{st}^{pq} (\mathbf q) = & - \sum\limits_{\mathbf k, l, m} a_l^{p\ast} (\mathbf k) a^t_l (\mathbf k)  a_m^{s\ast} (\mathbf k + \mathbf q) a_m^q (\mathbf k + \mathbf q) \\
& \times\frac{n_F (E_l (\mathbf k)) - n_F (E_m (\mathbf k + \mathbf q))}{E_l (\mathbf k) - E_m (\mathbf k + \mathbf q)} 
\end{aligned}
\label{eq:nonintsuscep}
\end{equation}
Matrix elements $a^t_{l} (\mathbf k)$ resulting from the diagonalization of the kinetic Hamiltonian $H_0$ connect orbital and band space denoted by indices $t$ and $l$ respectively. The energies $E_l$ are the eigenvalues of $H_0$ and $n_F(E)$ is the Fermi function. In our calculation both $\mathbf q$ and $\mathbf k$ run over uniform grids spanning the reciprocal unit cell. Temperature enters the calculation through the Fermi functions. 

The denominator of the fraction in Eq.~\ref{eq:nonintsuscep} can in principle vanish whenever the band energies 
$E_l$ and $E_m$ become equal. However, it can be shown easily using l'Hospital's rule that this does not lead to a diverging susceptibility.
\begin{equation}
\begin{split}
& \lim \limits_{E_l \to E_m} \frac{n_F(E_l (\mathbf k + \mathbf q)) - n_F(E_m (\mathbf 
k))}{E_l (\mathbf k + \mathbf q) - E_m (\mathbf k)} \\ 
= & - \beta \frac{e^{\beta  
E_l}}{(e^{\beta E_l} + 1)^2}
\end{split}
\label{eq:nonintsusceplhospital}
\end{equation}
Here, $\beta$ denotes the inverse temperature $\beta= (k_B T)^{-1}$. In practice we use this formula if the magnitude of the denominator falls below a certain 
threshold (e.g. $10^{-7}~{\rm eV}$). 

The observable non-interacting susceptibility is defined as the sum over all elements 
$\chi_{aa}^{bb}$ of the full tensor.
\begin{equation}
\chi (\mathbf q) = \frac{1}{2} \sum\limits_{a,b} \chi_{aa}^{bb} (\mathbf q)
\label{eq:nonintsuscepobservable}
\end{equation}

Within the random phase approximation the charge and spin susceptibilities can be calculated from the non-interacting susceptibility using inversion formulas involving the constant tensors $U^c$ and $U^s$ for the charge and spin channel, respectively.
\begin{subequations}
\begin{align}
\left[ (\chi^s (\mathbf q))_{st}^{pq} \right]^{-1} &= \left[ \chi_{st}^{pq} (\mathbf q) \right]^{-1} - (U^s)_{st}^{pq} \\[4pt]
\left[ (\chi^c)_{st}^{pq} (\mathbf q) \right]^{-1} &= \left[ \chi_{st}^{pq} (\mathbf q) \right]^{-1} + (U^c)_{st}^{pq}
\end{align}
\end{subequations}
Following Ref.~\onlinecite{Graser2009} the interaction tensors for the multi-orbital Hubbard model are given by
\begin{equation}
\begin{aligned}
( U^c )_{aa}^{aa} &= U & ( U^c )_{aa}^{bb} &= 2U' \\[2pt]
( U^c )_{ab}^{ab} &= \frac{3}{4} J -U' & ( U^c )_{ab}^{ba} &= J^\prime   \\[2pt]
( U^s )_{aa}^{aa} &= U & ( U^s )_{aa}^{bb} &= \frac{1}{2}J \\[2pt]
( U^s )_{ab}^{ab} &= \frac{1}{4} J +U' & ( U^s )_{ab}^{ba} &= J^\prime
.
\end{aligned}
\label{eq:interactiontensorcollected}
\end{equation}
The two-particle pairing vertex in the spin-singlet channel can be constructed from the charge and spin susceptibilities and the constant interaction tensors~\cite{Graser2009}.
\begin{equation}
\begin{aligned}
& \big( \Gamma^s \big) _{st}^{pq} (\mathbf k, \mathbf k^\prime) \\
= & \Big[  \frac{3}{2}  U^s   \chi^s   (\mathbf k \pm \mathbf k^\prime)  U^s  - \frac{1}{2}  U^c   \chi^c  (\mathbf k \pm \mathbf k^\prime)  U^c \Big]_{st}^{pq} \\
& +\frac{1}{2} \Big[  U^c  +  U^s  \Big]_{st}^{pq} 
\end{aligned}
\end{equation}
The previously calculated two-particle pairing vertex in orbital space is projected into band space using the matrix elements $a^t_{l} (\mathbf k)$ of the kinetic Hamiltonian.
\begin{equation}
\begin{aligned}
& \Gamma_{mn} (\mathbf k, \mathbf k^\prime) \\ 
= & \text{Re} \Bigg[ \sum\limits_{\substack{pq\\st}} a_m^{p \ast} (\mathbf k) a_m^{t \ast} (-\mathbf k) \, \big( \Gamma^s \big)_{st}^{pq}(\mathbf k, \mathbf k^\prime) \, a_n^{q} (\mathbf k^\prime) a_n^{s} (-\mathbf k^\prime) \Bigg]
\end{aligned}
\end{equation}
Restricting the momenta $\mathbf k$ and $\mathbf k^\prime$ in the pairing vertex to points $\mathbf k_m$ and $\mathbf k_n$ on the discretized Fermi surface, we can write down an effective eigenvalue equation.
\begin{equation}
\lambda g (\mathbf k_n) = - \sum\limits_{\mathbf k_m} \frac{\Gamma (\mathbf k_m, \mathbf k_n)}{\hbar |\mathbf v (\mathbf k_m)|} g (\mathbf k_m)
\end{equation}
Diagonalization of the kernel yields the symmetry eigenfunctions $g(\mathbf k_n)$ and corresponding eigenvalues $\lambda$, which characterize the strength of the electron pairing.

In our calculations the three-dimensional Fermi surface is discretized
using about 1000 $\mathbf k$-points. The susceptibility is calculated
on a $30 \times 30 \times 10$ $\mathbf k$-point grid at an inverse
temperature of $\beta = 40~{\rm eV}^{-1}$.

\section{Results}
\subsection{Electronic structure}
We investigated the electronic structure as a function of the lattice
parameter $a$ and the iron-selenium distance {\dse}. An overview of
electronic bandstructures as a function of the $a$ lattice parameter
with varied distance {\dse} is shown in Fig.~\ref{fig:bandsoverview},
while Fig.~\ref{fig:bandsheightcontrol} shows a scan of iron-selenium
distances {\dse} in a different range around the experimental value,
keeping constant the lattice parameter $a$. Corresponding Fermi
surfaces are shown in Figs.~\ref{fig:fsoverview} and
\ref{fig:fsheightcontrol}.

At small lattice parameters $a=0.93 a_{\rm exp}$ the Brillouin zone
comprises two hole pockets and two electron pockets
[Fig.~\ref{fig:bandsoverview} (a)-(c) and Fig.~\ref{fig:fsoverview}
(a)-(c)].  The iron-selenium distance {\dse} hardly influences the
electronic structure. An expansion of the iron lattice to $a=1.01
a_{\rm exp}$ decreases the size of the Fe $3d_{xy}$ hole pockets and
brings up a new Fe $3d_{xz / yz}$ hole pocket
[Fig.~\ref{fig:bandsoverview} (d)-(f) and Fig.~\ref{fig:fsoverview}
(d)-(f)].  The relative size is somewhat influenced by {\dse}. At
$a=4.1~\text{\AA}= 1.09a_{\rm exp}$ all Fermi surface pockets are
small. The relative position of Fe $3d_{xy}$ and Fe $3d_{z^2}$ bands
at the M=($\pi$,$\pi$,0) point is controlled by {\dse}. Around the experimental value
of {\dse} hole-pockets of Fe $3d_{z^2}$ character emerge around M
[Fig.~\ref{fig:bandsoverview} (h)-(i) and Fig.~\ref{fig:fsoverview}
(h)-(i)].

\begin{figure}[t]
\includegraphics[width=\linewidth]{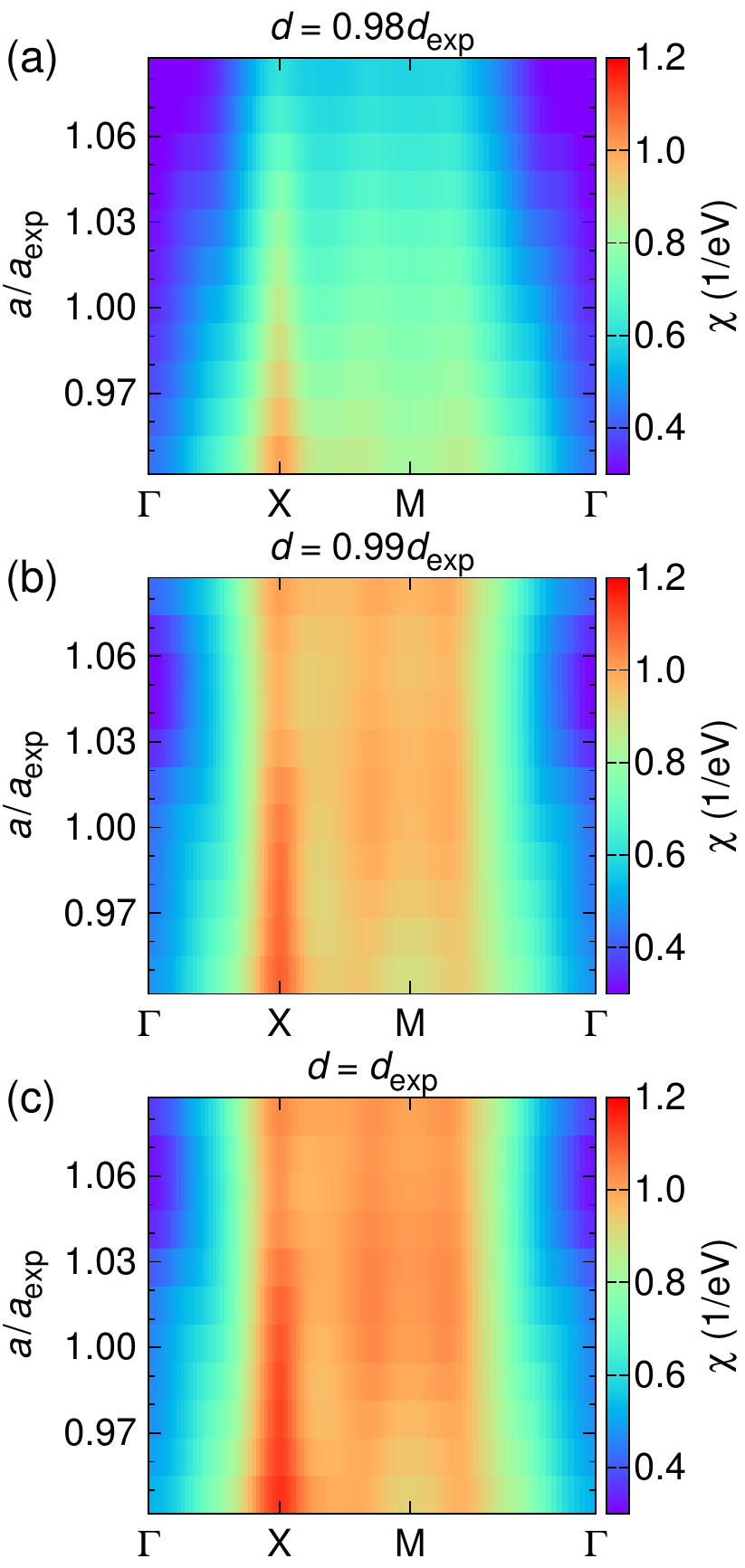}
\caption{(Color online) 
Static susceptibility on a high-symmetry path as a function of the lattice 
parameter $a$. The iron-selenium distance is kept fixed at (a) {\dse}=$0.98 d_{\rm Fe-Se~exp}$, (b) {\dse}=$0.99 d_{\rm Fe-Se~exp}$ and (c) {\dse}=$d_{\rm Fe-Se~exp}$.
}
\label{fig:astretchingsuscep}
\end{figure}

\begin{figure}[t]
\includegraphics[width=\linewidth]{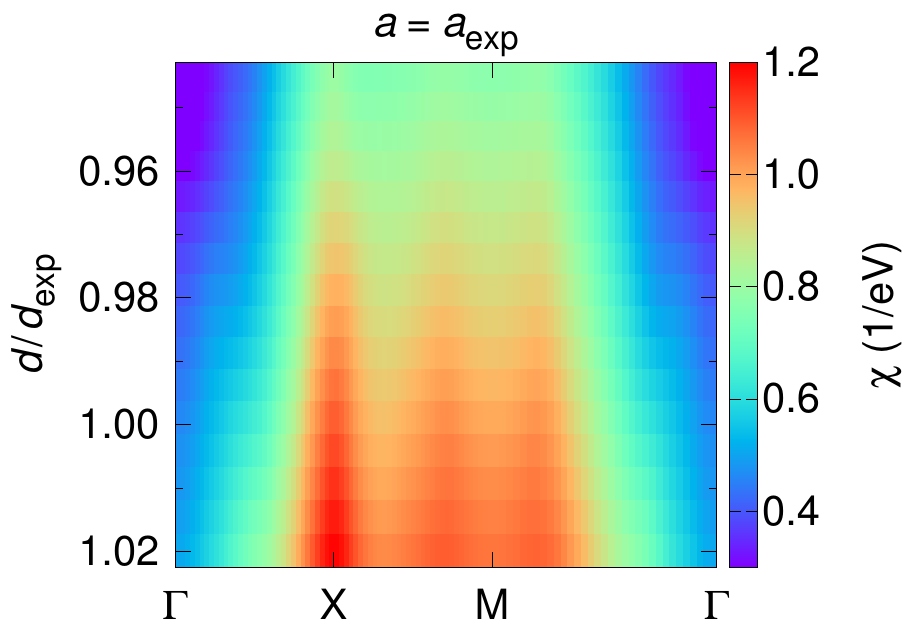}
\caption{(Color online) 
Static susceptibility on a high-symmetry path as a function of {\dse}
 at fixed lattice parameter $a=a_{\rm exp}$.
}
\label{fig:suscepheightcontrol}
\end{figure}

When tuning only {\dse} and keeping $a$ at the experimental value, the picture
is similar (Figs.~\ref{fig:bandsheightcontrol} and \ref{fig:fsheightcontrol}).
 As {\dse} decreases, the two present Fe $3d_{xz / yz}$ hole pockets
at $\Gamma$=(0,0,0) converge in size, while the electron pockets at X=($\pi$,0,0) shrink a little.
The Fe $3d_{xy}$ hole pockets around M finally dissappear, so that one ends up
with a Fermi surface with very small electron and hole pockets [see
Figs.~\ref{fig:bandsheightcontrol} (c) and \ref{fig:fsheightcontrol} (c)].

These observations can be directly related to the hopping parameters between Fe 
$3d_{xy}$ orbitals. While the next-nearest-neighbor hopping $t_2$ is roughly 
constant, the nearest-neighbor hopping $t_1$ is strongly influenced by the 
lattice expansion (see Fig.~\ref{fig:tbparameters}), which explains the strong 
influence of lattice parameters on the Fe $3d_{xy}$ hole pocket. Intuitively, 
one would expect that nearest-neighbor hopping $t_1$ increases upon compression,
while we find a strong decrease. 

The reason for this behavior has been exhaustively assessed in 
Ref.~\onlinecite{Suzuki2014} (see in particular Fig.~4 therein). In summary, the 
nearest-neighbor hopping $t_1$ is the sum of direct Fe-Fe hopping paths and 
indirect Fe-Se-Fe paths. Direct and indirect contributions to $t_1$ have 
opposite signs. At equilibrium lattice parameters the indirect hopping contributions 
to $t_1$ dominate. Therefore, as direct Fe-Fe hopping is enhanced upon compression, $t_1$ 
counterintuitively decreases and even changes sign in the extreme compressed 
regime [see Fig.~\ref{fig:tbparameters}~(a-c)]. The next-nearest neighbor 
hopping $t_2$, however, is hardly affected by compression, since it depends in first approximation only on the Fe-Se-Fe hopping. This picture is clearly consistent also 
with Fig.~\ref{fig:tbparameters}~(d), where $t_1$ decreases as the Fe-Se 
distance is increased. In this case direct Fe-Fe hoppings stay constant, while 
indirect hopping via Se decreases due to the larger Fe-Se distance, so that $t_1$ finally vanishes in the most 
expanded case we investigated. The importance 
of the balance of $t_1$ and $t_2$ for superconductivity in iron-based materials 
has been assessed in previous publications~\cite{Suzuki2014, Guterding2015a}.

\subsection{Non-interacting observable susceptibility}
The non-interacting susceptibility is calculated from
Eqs.~\ref{eq:nonintsuscep} and \ref{eq:nonintsuscepobservable}. The
static susceptibility on the high-symmetry path $\Gamma$-X-M-$\Gamma$
is shown in Fig.~\ref{fig:astretchingsuscep} as a function of the $a$
lattice parameter and {\dse}. In Fig.~\ref{fig:suscepheightcontrol}
the static susceptibility is shown as a function of {\dse} in a
somewhat different range, but at constant lattice parameter $a_{\rm
  exp}$.

A decrease of {\dse} at constant $a_{\rm exp}$ weakens the
susceptibility (Fig.~\ref{fig:suscepheightcontrol}), while an increase 
in {\dse} strongly enhances the susceptibility.
A change
in the lattice parameter $a$ (Fig.~\ref{fig:astretchingsuscep})
produces more interesting behavior: Except at {\dse}$=0.98 d_{\rm
  {Fe-Se}~exp}$ only the susceptibility at the X point is selectively
weakened by the lattice expansion, while the susceptibility around the
M point is enhanced. Since the static susceptibility at the X point is
associated with stripe antiferromagnetism, while the static
susceptibility at M is associated with checkerboard
antiferromagnetism, this nontrivial behavior of the static
susceptibility might be important for extremely stretched films of
iron selenide, where stripe AFM order can be expected to compete with
checkerboard AFM order.~\cite{Tresca2015a}

\begin{figure}[t]
  \includegraphics[width=\linewidth]{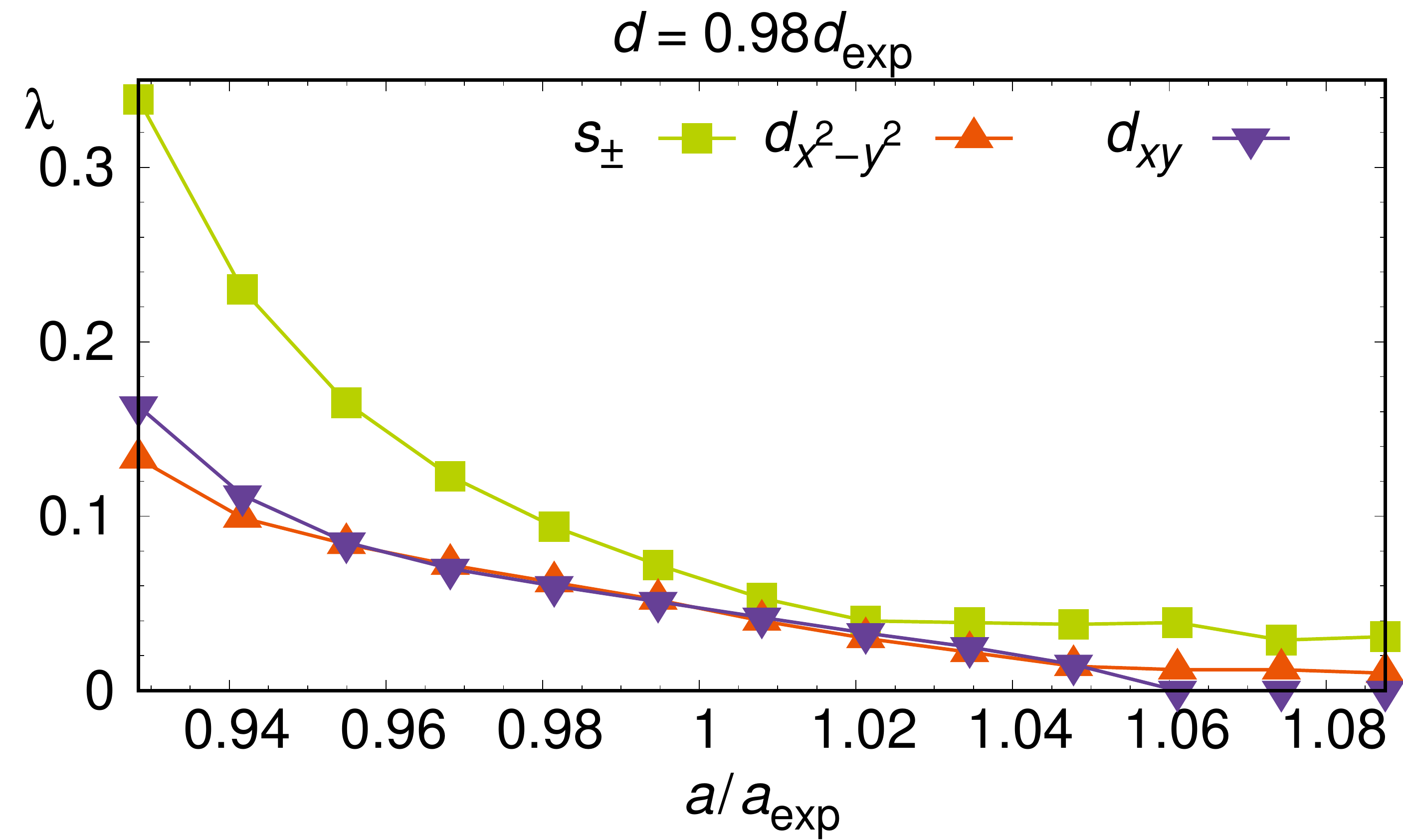}\caption{(Color
    online) Pairing eigenvalues $\lambda$ as a function of the lattice
    parameter $a$ at fixed iron-selenium distance $d=0.98d_{\rm
      Fe-Se~exp}$.  }
\label{fig:pairingd2p35}
\end{figure}

In the following we explain the behavior of the static susceptibility based on the evolution of 
the electronic bandstructures and Fermi surfaces. The monotonous decrease of the susceptibility for constant lattice 
parameter $a$ and varied {\dse} (Fig.~\ref{fig:suscepheightcontrol}) is due to the vanishing of 
the Fe $3d_{xy}$ hole pocket at the M point as shown in 
Figs.~\ref{fig:bandsheightcontrol} and \ref{fig:fsheightcontrol}. The shape of 
the Fermi surface otherwise remains unaltered, so that only a quantitative 
decrease of the susceptibility is observed due to the reduced spectral weight at 
the Fermi level. The same is true for varied lattice parameter $a$ and {\dse}$=0.98d_{\rm Fe-Se~exp}$ [compare Figs.~\ref{fig:bandsoverview} (a,d,g) to Fig.~\ref{fig:bandsheightcontrol} and Figs.~\ref{fig:fsoverview} (a,d,g) to Fig.~\ref{fig:fsheightcontrol}]. Likewise Figs.~\ref{fig:tbparameters} (a) and (d) contain similar trends for the hopping parameters between Fe $3d_{xy}$ orbitals.

\begin{figure}[t]
\includegraphics[width=\linewidth]{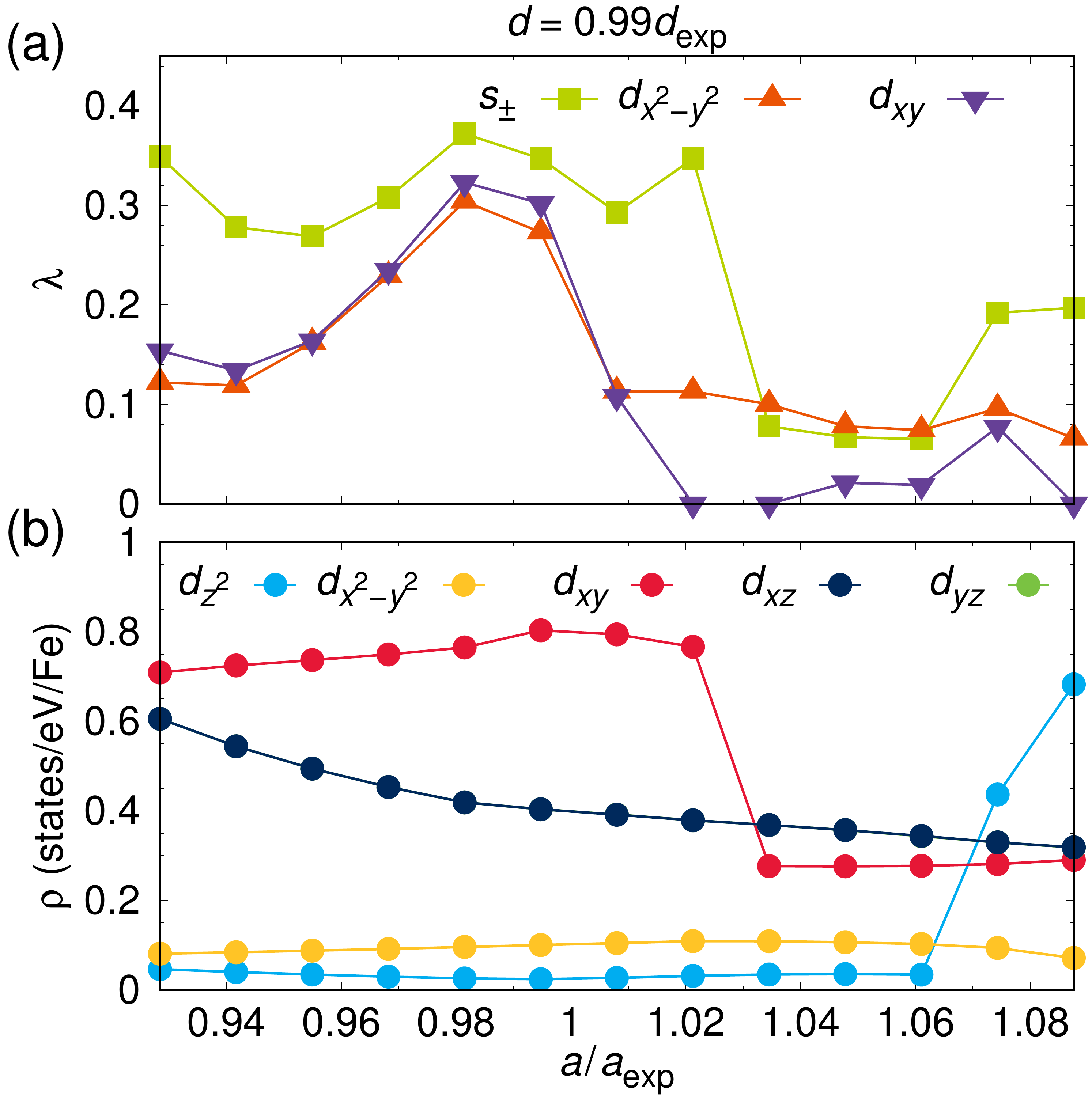}\caption{(Color online) (a) Pairing eigenvalues $\lambda$ and (b) orbital-resolved density of states $\rho$ at the Fermi level as a function of the lattice parameter $a$ at fixed iron-selenium distance $d=0.99d_{\rm Fe-Se~exp}$.
}
\label{fig:pairingd2p37}
\end{figure}

Furthermore, while
 the reduction of the susceptibility at the X point is 
obvious for constant {\dse} at varied lattice 
parameters $a$ (see Fig.~\ref{fig:astretchingsuscep}), the increased
susceptibility at the M point is the outstanding non-trivial feature upon lattice
expansion. 
Also here we observe the Fe $3d_{xy}$ hole pocket vanish, 
but the electron pockets located around X and the hole pockets in the Brillouin zone center contract notably [see Figs.~\ref{fig:bandsoverview} (h-i) and \ref{fig:fsoverview} (h-i)]. This means that the increased susceptibility at M is due to enhanced scattering processes involving the contracted Fe $3d_{xy}$ electron pockets. This is also made plausible by the location of maxima in susceptibility in Fig.~\ref{fig:astretchingsuscep}, which seem to converge towards the M point with enlarged lattice parameter 
$a$, i.e. when the electron pockets contract.

\subsection{Symmetry and strength of electron pairing}
\begin{figure}[t]
\includegraphics[width=\linewidth]{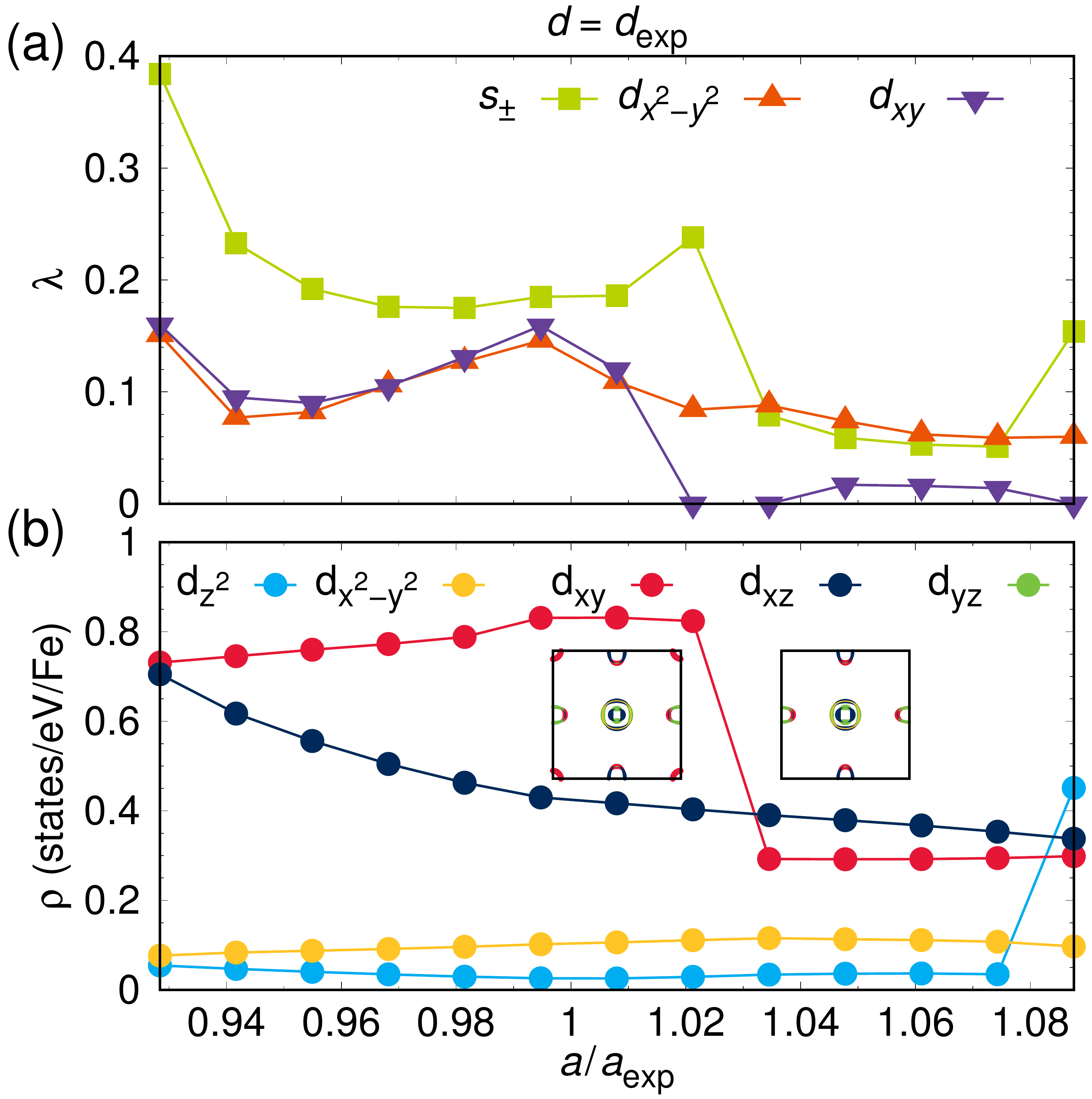}\caption{(Color
online) (a) Pairing eigenvalues $\lambda$ and (b) orbital-resolved density of
states $\rho$ at the Fermi level as a function of the lattice parameter $a$ at
fixed iron-selenium distance {\dse}$=d_{\rm Fe-Se~exp}$.  }
\label{fig:pairingd2p393}
\end{figure}

\begin{figure}[t]
\includegraphics[width=\linewidth]{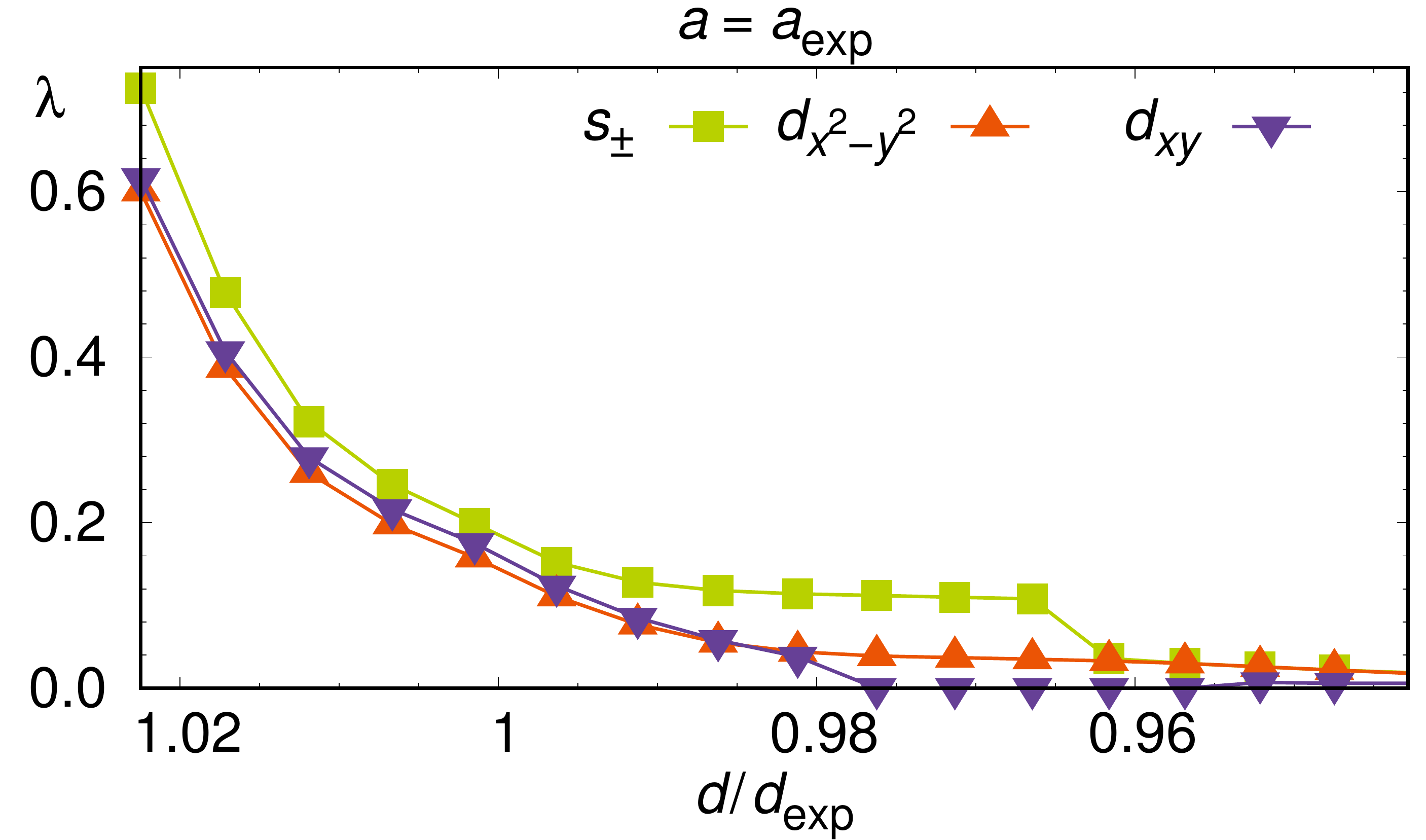}\caption{(Color
online) Pairing eigenvalues $\lambda$  as a function of the iron-selenium
distance {\dse} at fixed lattice parameter
 $a=a_{\rm exp}$.  }
\label{fig:pairingheightcontrol}
\end{figure}

\begin{figure}[t]
\includegraphics[width=\linewidth]{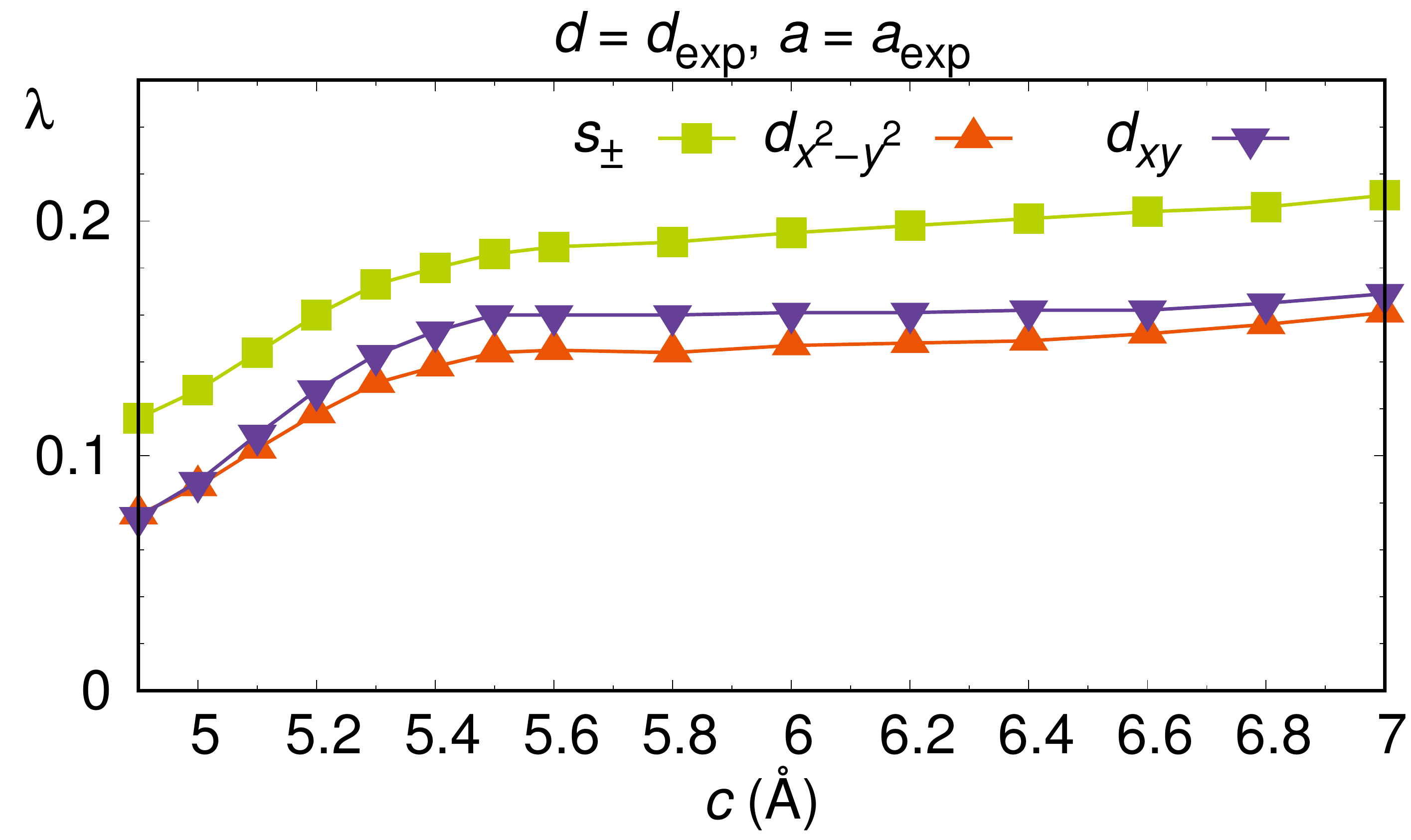}\caption{(Color
online) Pairing eigenvalues $\lambda$  as a function of the lattice parameter
$c$ at fixed lattice parameter $a= a_{\rm exp}$ and fixed
iron-selenium distance {\dse}=$d_{\rm Fe-Se~exp}$.}
\label{fig:pairingcaxiscontrol}
\end{figure}

Several scans of the pairing strength and symmetries are shown in
Figs.~\ref{fig:pairingd2p35}, \ref{fig:pairingd2p37},
\ref{fig:pairingd2p393}, \ref{fig:pairingheightcontrol} and \ref{fig:pairingcaxiscontrol}.
The pairing symmetries with largest eigenvalues can be characterized as $s_\pm$ (sign-changing $s$-wave), $d_{x^2-y^2}$ and $d_{xy}$. These are the symmetry eigenfunctions that usually compete in iron-based superconductors.

For small {\dse}=$0.98 d_{\rm Fe-Se~exp}$ at varied $a$ and for constant $a$ at varied {\dse} the results do not contain much structure. The pairing eigenvalues monotonously decrease with increased $a$ and with decreased {\dse} (see Figs.~\ref{fig:pairingd2p35} and \ref{fig:pairingheightcontrol}). 
For larger {\dse}, the pairing strength and the symmetry of the leading
solution in the pairing eigenproblem on the
lattice parameter $a$ mirrors the non-trivial results we obtained for the susceptibility (see Figs.~\ref{fig:pairingd2p37} and
\ref{fig:pairingd2p393}).

For varied lattice parameter $a$ and constant iron-selenium distance
{\dse}$\geq 0.99 d_{\rm Fe-Se~exp}$ we observe a breakdown of the $s_\pm$
solution (see Figs.~\ref{fig:pairingd2p37} and
\ref{fig:pairingd2p393}) at the Lifshitz transition, where the Fe
$3d_{xy}$ hole pocket disappears, as clearly indicated by the jump in
the Fe $3d_{xy}$ density of states at the Fermi level (see also
Fig.~\ref{fig:fsoverview}). Just before the breakdown, the $s_\pm$
pairing is quite strongly enhanced, probably because of the enhanced
density of states at the Fermi level contributed by the upper edge of
the hole band.

Before the breakdown of $s_\pm$ pairing, the $d_{xy}$ pairing state is
already completely suppressed. This is the case because a nodal
$s_\pm$ state can also accommodate susceptibilities with wave vector
around the M point via the introduction of nodes into the order
parameter, while the inclusion of additional nodes is very hard for
the $d_{xy}$ state. However, the $d_{x^2-y^2}$ state takes the lead
when the Fe $3d_{xy}$ hole pocket has disappeared. This fact relates
directly to the emergent peak in the susceptibility at the M point and
the strong decrease at the X point. Note, however, that the pairing
strength is severely decreased after the Lifshitz transition.

In the intermediate region there is strong competition between
$d_{xy}$ and $d_{x^2-y^2}$ pairing for the first subleading solution,
while $s_\pm$ remains the leading solution. The intermediate
enhancement of $d_{xy}$ pairing occurs whenever the geometry of the Fe
$3d_{xz/yz}$ hole and electron pockets matches well. In this case, the
$s_\pm$ pairing state can, however, avoid nodes on the Fermi surface
altogether and consequently takes the lead. Platt {\it et al.} explain
this general phenomenon in terms of optimization of the condensation
energy~\cite{Platt2013}.

Remarkably, in Fig~\ref{fig:pairingheightcontrol} an expansion of {\dse} leads 
to a similarly strong increase of pairing eigenvalues regardless of the 
associated symmetry eigenfunction. This is a consequence of the overall strongly 
enhanced susceptibility we observed in Fig.~\ref{fig:suscepheightcontrol}, which 
does have a maximum at the X point, but less well separated from other peaks 
than in Fig.~\ref{fig:astretchingsuscep}. Therefore, it is likely that a strong 
degeneracy of different magnetic states is present under expansion of {\dse}. 
The presence of magnetic degeneracy should suppress possible competing long 
range  magnetic order and is likely beneficial for superconductivity if the 
degeneracy is between pairing-promoting states. To elucidate this issue, DFT 
calculations for different magnetic states could be instructive, but are beyond 
the scope of the present paper. 

A scan of the dependence on the $c$ lattice parameter is shown in
Fig.~\ref{fig:pairingcaxiscontrol}, which is reminiscent of the plot
by Noji \textit{et al.}~\cite{Noji2014} that showed a levelling-off of
the superconducting transition temperatur $T_c$ in certain
intercalates when the distance between FeSe layers is
increased. However, the critical length is somewhat smaller here than
in typical intercalates, probably because in our model study presented
here we do not actually intercalate anything in between of the FeSe
layers, but leave the space in between as vacuum. Note that the actual
behavior of those intercalates is not only a function of layer
distance, but also crucially depends on electron
doping~\cite{Guterding2015a, Hayashi2015, Sun2015}.

As a final note, FeS~\cite{Lai2015} has lattice parameters $a =
3.6802~{\rm \AA}$ and $d_{\rm Fe-S} = 2.235~{\rm \AA}$ with
relations relative to the experimental values of FeSe, $a_{\rm FeS} /
a_{\rm FeSe} = 0.98$ and $d_{\rm Fe-S} / d_{\rm Fe-Se} = 0.93$.  This corresponds
to an extremely compressed case, further reducing the $a$ parameter in
Figs.~\ref{fig:bandsheightcontrol}~(c) and~\ref{fig:fsheightcontrol}
~(c) (see Ref.~\onlinecite{Tresca2017}). Extrapolating the trends of
Fig.~\ref{fig:pairingheightcontrol} (a) and
Fig.~\ref{fig:pairingcaxiscontrol} it is to be expected that the
susceptibility for this case will be rather featureless which would
further reduce $T_c$. Actually, $T_c$ in FeS is 4~K.

\section{Summary}
We have discussed that the underlying electronic structure of iron selenide depends both on the lattice spacing and the  iron-selenium distance in a nontrivial way, leading to complex behavior of non-interacting susceptibility and electron pairing, with several competing channels. Our main observations are: (i) Expansion of the Fe square lattice parameter at constant Fe-Se distance should
enhance $T_c$ up to a point before a switch of superconducting order
parameter from $s_\pm$ to $d_{x^2-y^2}$ occurs. (ii) Increasing Fe-Se
distance at constant Fe square lattice parameter $a$ should enhance
$T_c$.  (iii) Increasing the $c$ lattice parameter at constant
experimental {\dfe} and {\dse} distances (i.e. increasing the van der
Waals gap) slightly increases $T_c$.  (iv) Compression of the Fe-Fe
square lattice at compressed Fe-Se distance significantly enhances
$T_c$.  Observation (iii) is essentially known from charge neutral
FeSe intercalatates. Observation (iv) is very consistent with the well
known pressure enhancement of $T_c$~\cite{Mizuguchi2008}.
 However, observations (i) and
(ii) could lead to new design ideas.

\begin{acknowledgments}
The authors acknowledge support from the German Research Foundation (Deutsche Forschungsgemeinschaft). Furthermore, the authors would like to thank Matthew D. Watson, Amalia I. Coldea, Steffen Backes, Andreas Kreisel and Peter J. Hirschfeld for helpful discussions.
\end{acknowledgments}

\end{document}